\documentclass[11pt,twoside]{article}
\usepackage{asp2010}

\newcommand{\msolar}{\mbox{\,$M_{\odot}$}}

\newcommand{\ovi}{{\rm O{\textsc{vi}}\,}}
\newcommand{\ovii}{{\rm O{\textsc{vii}}\,}}

\newcommand{\novi}{{\rm N(O{\textsc{vi}})\,}}

\resetcounters

\begin{document}

\title{A Non-Equilibrium Ionization Model of the Local and Loop I Bubbles - Tracing the \ovi Distribution}
\author{Miguel A. de Avillez$^{1}$, Dieter Breitschwerdt$^{2}$, Emanuele Spitoni$^{1}$ \& Nuno
Carvalho$^{1}$
\affil{$^{1}$Dept. of Mathematics, U. \'Evora, R. Rom\~ao Ramalho 59, \'Evora, Portugal}
\affil{$^2$ZAA, Technische Universit\"at Berlin, Hardenbergstr.~36, Berlin, Germany}
}

\begin{abstract}
We present the first to date three-dimensional high-resolution hydrodynamical simulation 
tracing the non-equilibrium ionization evolution (using the Eborae Atomic and Molecular Plasma
Emission Code - E(A+M)PEC) of the Local Bubble and Loop I bubbles embedded in a turbulent
supernova-driven interstellar medium.
\end{abstract}

\section{Introduction}
The Local Bubble (LB), hosting the Local Cloud surrounding the solar system, is an X-ray emitting region
extending 100 pc in radius, and it is embedded in a somewhat larger H{\sc i} deficient cavity.
\begin{figure}[thbp]
\centerline{\includegraphics[width=0.65\hsize,angle=0]{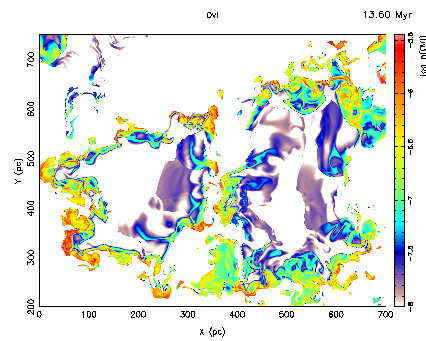}}
\caption{\ovi density distributions in the LB (centered at ($x=250, y=450$) pc) and Loop I (centered at
($x=480, y=400$) pc) at 0.5 Myr after the last SN in the LB, which occurred at evolution time 13.1 Myr. Both
bubbles are surrounded by thin fragmented \ovi shells.}
\label{ovidist}
\end{figure}
Its origin and spectral properties in UV, EUV and X-rays are still poorly understood. Standard LB models fail
to reproduce the observed low OVI absorption column density. Heliospheric in situ measurements are sensitive
to the boundary conditions imposed by the LB and the OVI column density in absorption is a crucial test for
modelling of the local ISM. We investigate if in the multisupernova scenario \citep{fu2006} the observed O{\sc
vi} column density in absorption (along lines o sight (LOS) crossing the LB) can be reproduced. 

\section{Model and Simulations}

We use the 3D supernova-driven ISM model of \citet{ab2009} with new features: (i) Simultaneous evolution of
the Local and Loop~I superbubbles as a result of the successive explosions of massive stars from a moving
subgroup - 17 stars with masses $\in[21.5,8.2]$ \msolar\  and Sco Cen - 39 stars with masses $\in[14,31]$
\msolar\ clusters \citep{fu2006,egg1998}, and (ii) \emph{Time-dependent evolution} of the ionization structure
of H, He, C, N, O, Ne, Mg, Si, S and Fe ions with latest solar abundances \citep{agss2009} using
E(A+M)PEC (see Avillez \& Spitoni in this book).

\section{Results and Final Remarks}

The locally enhanced SN rates produce coherent structures within a highly disturbed background medium. The
Local and Loop I bubbles fill volumes roughly corresponding to the present day observations
(Figure~\ref{ovidist}). The \ovi distribution inside the LB has been traced by column density measurements
through LOS taken from the Sun's vantage point, located at $(x=250, y=450)$ pc and 90 pc from the interaction
region between the Local and Loop~I bubbles (Figure~\ref{ovidist}). The main results of these LOS observations
are: (1) \novi in the simulated bubble grows with time as a result of \ovii recombination (with 
delayed recombination playing a role) reaching the \novi values observed with FUSE; (2) Only for
$0.6<\Delta$t$_{SN}\leq 0.9$ Myr (since the last SN occurrence in the cavity) the simulated
average and maximum \novi are within the minimum and maximum observed column densities by FUSE
\citep{oe2005,sl2006,bow2008}; (3) The number of lines of sight with $10^{12} <\novi < 10^{13} \, {\rm
cm}^{-2}$ increase with time towards 88\% at $\Delta$t$_{SN}=0.9$ Myr since the last SN occurrence; (4) a
fragmenting LB shell is consistent with spectral variations in the ROSAT  R1 and R2 bands
\citep{bfe2000}.

This work strengthens the \emph{importance of taking into account all the relevant atomic processes within a
self-consistent evolutionary picture} of the Local Bubble in particular, and of the interstellar medium in
general.

\acknowledgements 

Research funded by FCT project PTDC/CTE-AST/70877/2006.

\bibliography{author}

\end{document}